\documentclass[twocolumn, prl]{revtex4}
\pdfoutput=1
\usepackage{graphicx}
\usepackage{amsmath}
\usepackage{epsfig}
\begin{document}
\title{ A Two-dimensional Algebraic Quantum Liquid Produced by an Atomic Simulator of the Quantum Lifshitz Model}
\author{ Hoi Chun Po$^{1,2}$, Qi Zhou$^1$}
\affiliation{$1$ Department of Physics, The Chinese University of Hong Kong, Shatin, New Territories, Hong Kong\\
$2$  Department of Physics, University of California, Berkeley, California 94720, USA}
\date{\today}

\begin{abstract}

Bosons have a natural instinct to condense at zero temperature.  It is a long-standing challenge to create a high-dimensional quantum liquid that does not exhibit long-range order at the ground state, as either extreme experimental parameters or sophisticated designs of microscopic Hamiltonian are required for suppressing the condensation. Here, we show that ultra cold atoms with synthetic spin-orbit coupling provide physicists a simple and practical scheme to produce a two-dimensional algebraic quantum liquid at the ground state. This quantum liquid arises at a critical Lifshitz point, where the single-particle ground state shrinks to a point from a circle in the momentum space, and many fundamental properties of two-dimensional bosons are changed in its proximity. Such an ideal simulator of the quantum Lifshitz model allows experimentalists to directly visualize and explore the deconfinement transition of topological excitations, an intriguing phenomenon that is difficult to access in other systems. 
\end{abstract}

\maketitle

Bosons are well known for preferring to form a Bose-Einstein condensate (BEC) at low temperatures. Such is the case for most bosonic systems in three dimensions. In lower dimensions, the reduced coordination enhances quantum fluctuation and BEC is either absent (one dimension) or confined to strictly zero temperature (two dimensions). Whereas these textbook results of the ground state of bosons are intrinsically determined by the fundamental  Bose-Einstein statistics and can be qualitatively understood in the non-interacting limit, there have been intensive interests to explore schemes for suppressing condensation at zero temperature. The success of such an effort will pave the way for creating novel quantum body ground states without ordering\cite{Rot1, Rot2, Rot3, Rot4, Xu1, Xu2, Xu3, Fisher1, Fisher2, Fisher3}. However, a challenge is that the currently devised  schemes require either extreme experimental conditions, such as a fast rotation of an atomic cloud at a frequency extremely close to that of the trapping potential,  or delicate designs of sophisticated Hamiltonian, such as lattice models containing ring exchange or even more complicated terms. The lofty goal of experimentally realizing a non-condensed quantum liquid as the ground state in high dimensions has not been achieved yet. 

Synthetic spin-orbit coupling is one of the most important developments in current studies of ultra cold atom physics\cite{Ian,  Zhang, Martin, Chen, ChenY, Engels}. Whereas the current interest on this topic has been mainly focusing on topological matters, here we show that synthetic spin-orbit coupling provides one an unprecedented means to suppress the condensation in two dimensions even at zero temperature and produce an algebraic quantum liquid as the ground state of interacting bosons. This novel  quantum state is induced by a fully quartic dispersion at a Lifshitz point, where the quadratic term of the spatial gradient vanishes in the Hamiltonian describing low energy physics. Surprisingly,  such a synthetic spin-orbit coupling leads to a natural realization of quantum Lifshitz model, an important theoretical tool for studying a wide range of exotic phenomena in modern physics, including the deconfinement transition in condensed matter physics\cite{Fradkin, Kivelson, dimer, Moessner1, Fradkin2, Ashvin} and quantum gravity in high energy physics \cite{Petr, Petr2, Holsheimer, Charmousis}.  Despite its profound applications in fundamental physics, such a model has not been realized in a realistic system before, because of the difficulty of suppressing the quadratic term of the spatial gradient that is always dominant in the low-energy effective theories for ordinary quantum many-body systems. Our work shows that the high controllability of ultra cold atoms allows one to create an ideal simulator of quantum Lifshitz model, which leads to a controllable scheme to access the Lifshitz point, where exotic quantum phenomena occur. This atomic simulator can be used to directly probe intriguing phenomena, including vanishing Berezinskii-Kosterlitz-Thouless (BKT) transition temperature and the deconfinement transition of vortices  in superfluids. \\

 \begin{figure*}[tbp]
\begin{center}
\includegraphics[width=6in]{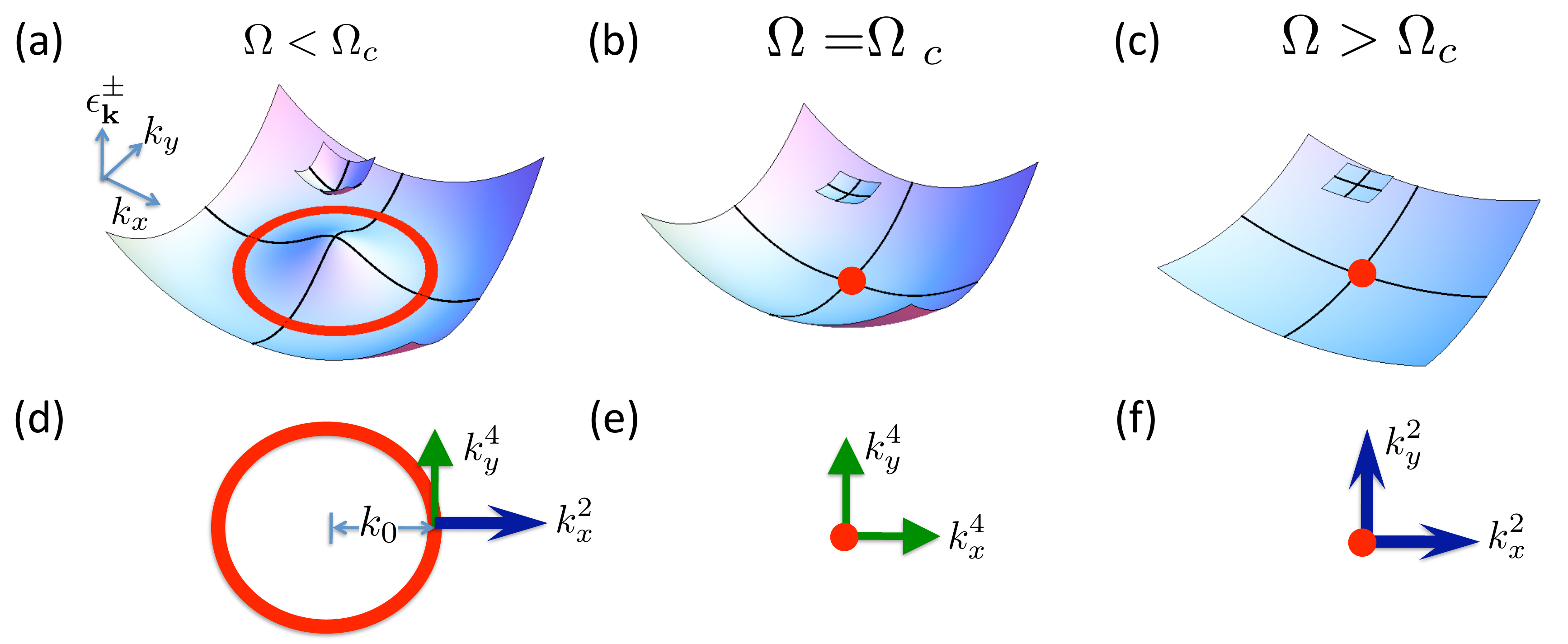}
\end{center}
\caption{Single-particle dispersion $\epsilon_{\bf k}^{\pm}$ at different values of $\Omega$.  (a) The minimum of $\epsilon_{\bf k}^{-}$ forms a circle if $\Omega<\Omega_c$. $\epsilon_{\bf k}^{-}$ is quadratic and quartic along the radial and tangent direction of the circle respectively. (b) The circle shrinks to a single point at ${\bf k}=0$ when $\Omega=\Omega_c$. At this critical point, $\epsilon_{\bf k}^{-}$ becomes quartic at small $|{\bf k}|$. (c) If $\Omega>\Omega_c$, the minimum remains to be a single point in the momentum space, and $\epsilon_{\bf k}^{-}$ becomes quadratic again at small $|{\bf k}|$. (d-f) Top view of the single-particle energy minimum. Blue long and green short arrows represent the quadratic and quartic dispersions along the $k_x$ and $k_y$ directions. }
 \end{figure*}

{\bf Quartic dispersion and effective dimension reduction}  We consider the Hamiltonian of two-dimensional bosons with spin-orbit coupling (SOC),
\begin{equation}
\mathcal{K}=\frac{1}{2m}\left(\hat{p}_x^2+\hat{p}_y^2-2\lambda(\sigma_x\hat{p}_x+\eta\sigma_y \hat{p}_y)+2\lambda\Omega\sigma_z\right), \label{Ki}
\end{equation}
where $m$ is the mass,  $\hat{p}_{i=x,y}$ is the momentum operator, $\lambda$ and $0\le \eta\le 1$ characterize the strength and anisotropy of SOC repsectively. $\hbar$ and $k_b$ are been set to be 1 in this manuscript.  For $\eta=0$, $\mathcal{K}$ describes the synthetic SOC produced by the Raman scheme, where $\Omega$ is proportional to Raman frequency. For $\eta=1$,  $\mathcal{K}$ is identical  to Rashba coupling with a magnetic field along the $z$ axis. Many theoretical studies have proposed how to produce a SOC with finite strengths along multiple spatial directions\cite{Dalibard, Anderson2,  Liu}. It is very promising that a fully controllable Hamiltonian in equation (\ref{Ki}) will be realized in the near future. 

We start from the isotropic case where $\eta=1$. For fermions, this model has been extensively studied in the context of topological matters. For bosons, it has been much less explored except for the special case with $\Omega=0$. Including a finite $\Omega$, the kinetic energy can be written as
\begin{equation}
\epsilon_{\bf k}^{\pm}=\frac{1}{2m}\left(k^2\pm 2 \lambda\sqrt{\Omega^2+k^2}\right),
\end{equation}
where $\pm$ corresponds to the upper and lower branch respectively. It is straightforward to show that there exists a critical value $\Omega_c^0=\lambda$. If $\Omega>\Omega_c^0$, $\epsilon_{\bf k}^{\pm}$ has a unique minimum at ${\bf k}=0$.  When $\Omega<\Omega_c^0$, the minimum forms an infinitely degenerate circle with a radius $|{\bf k}|\equiv k_0=\sqrt{\lambda^2-\Omega^2}$.  This circle shrinks to a single point at ${\bf k}=0$ when $\Omega=\Omega_c^0$, and $\epsilon_{\bf k}^-$ at small $k=|{\bf k}|$ may be expanded as
\begin{equation}
\epsilon_{\bf k}^-=\frac{1}{2m}\left(-2\lambda^2+\frac{1}{4\lambda^2}k^4+O(k^6)\right), \label{kn}
\end{equation}
which becomes  quartic other than the conventional quadratic ones of ordinary particles, as shown in figure (1).  If one considers the identity, $\int d{\bf k}=\int d\epsilon \mathcal{N}(\epsilon)$, where $\mathcal{N}(\epsilon)$ is the density of states and $\epsilon\sim |{\bf k}|^4$ at small $|{\bf k}|$, one immediately sees  that $\mathcal{N}(\epsilon)$ at low energies $\epsilon\rightarrow 0$ becomes $\mathcal{N}(\epsilon)\sim \epsilon^{-1/2}$, similar to the one of an ordinary one-dimensional system. Without physically reducing the dimension of the system, for instance, by imposing a strong confinement potential to completely quench the kinetics along one spatial direction, spin-orbit coupling here partially suppresses the kinetic energy along all spatial directions through changing the ordinary quadratic dispersion to a much flattened one $\sim |{\bf k}|^4$, and leads to an effective dimension reduction. As seen from the density of states, such an effective dimension reduction allows one to directly conclude that for non-interacting bosons, a condensate is absent even at zero temperature when $\Omega=\Omega_c^0$. If one includes interaction effects, as shown later, such a fully quartic dispersion along all the spatial directions is the microscopic origin for the rise of a Lifshitz point in the low-energy effective theory\cite{Lifshitz, Hornreich}. It is worth pointing out that a similar quartic dispersion in two dimensions can be produced in a shaken square lattice (Supplementary Material).





\begin{figure*}[tbp]
\begin{center}
\includegraphics[width=7in]{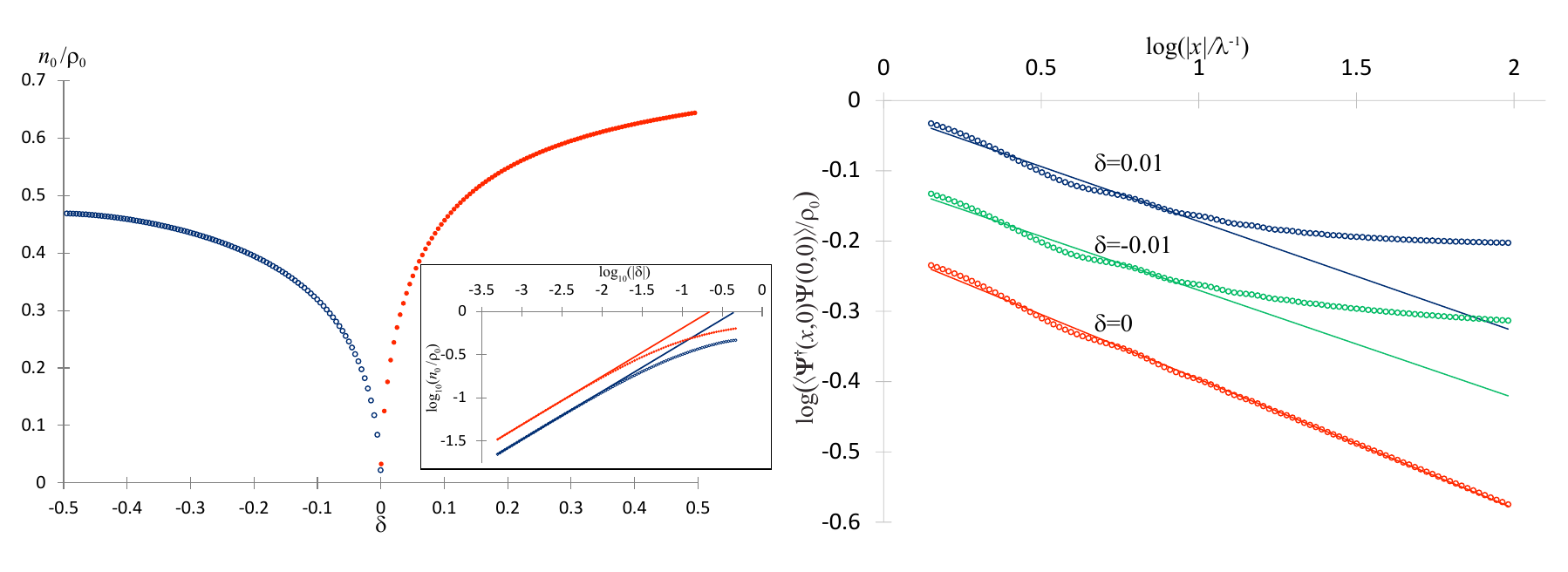}
\end{center}
\caption{Condensate fraction and correlation function. Left: Condensate fraction as a function of $\delta=(\Omega-\Omega_c)/\Omega_c$ evaluated using parameters $\rho_0 = 0.125 \lambda^2, g_0 = 0.8 m^{-1}, g= -0.001 g_0$. At the critical point $\delta=0$, condensate fraction is suppressed down to zero. Near the critical point, it scales as a power-law function of $\delta$ on both sides of the critical point. Inset is the log-log plot, which directly shows the power-law scaling when $\delta\rightarrow 0$. Right: a log-log plot of correlation function at different $\delta$ evaluated using parameters $\rho_0 = 1.25 \lambda^2, g_0 = 0.8 m^{-1}, g= -0.001 g_0$. At the critical point, the correlation function is completely a power-law function. The linear fit provides $K_c=2.72$, close to the analytical result $2.78$. Near the critical point, the correlation function approaches a constant when $|x|\rightarrow \infty$, and the power-law feature emerges in an intermediate length scale.  }
 \end{figure*}

{\bf A simulator of Quantum Lifshitz model }  Ultra cold bosons interact through a contact potential,  
\begin{equation}
\mathcal{U}=\int d{\bf r}\left( \left( g_0+\frac{g_s}{2}\right)\hat{n}^2-\frac{g_s}{2}\hat{S}_z^2\right)\label{U},
\end{equation}
where $\hat{n}=\hat{n}_\uparrow+\hat{n}_\downarrow$ is the total density operator, and $\hat{S}_z=\hat{n}_\uparrow-\hat{n}_\downarrow$. $g_0+\frac{g_s}{2}$ and $-\frac{g_s}{2}$ characterize the strength of density-density interaction and spin-dependent interaction respectively.  Mean field solutions are pursued by minimizing the ground state energy using the ansatz 
\begin{equation}
\Psi=e^{\frac{i\sigma_y\pi}{4}}\sqrt{\rho}e^{i\theta}\left(\begin{array}{c}
-\sin(\phi/2)e^{-i\chi/2}\\
\cos(\phi/2)e^{i\chi/2}
\end{array}\right), \label{an}
\end{equation}
where $\rho$ and $\theta$ are the density and the phase respectively. $\phi$ and $\chi$ characterize the spin orientation. A unitary transformation $U=e^{\frac{i\sigma_y\pi}{4}}$ is introduced for avoiding the ambiguity of defining $\chi$ at the north and south poles of the Bloch sphere.   

Mean field results show that interaction shifts the critical point to $\Omega_c=\lambda-g_s m\rho/\lambda$. $\Omega_c$ reduces to $\Omega_c^0$ if $g_s=0$. The mean field value of the phase $\theta_0$ is $0$ and $\lambda \sqrt{1 - \Omega^2 /\Omega_c^2}\, x$ for $\Omega>\Omega_c$ and $\Omega<\Omega_c$ respectively, analogous to the zero-momentum\cite{Li} and plane-wave condensate\cite{Zhai, Ho} in three dimensions. Whereas both states break $U(1)$ symmetry, the latter one also breaks the rotation symmetry in the momentum space, as interaction lifts the infinite degeneracy on the circle of kinetic energy minimum\cite{Zhou}. Another (first order) transition from the plane-wave phase to a stripe phase at $\Omega_c'<\Omega_c$ is not relevant to our discussions here. To include quantum fluctuations, we introduce $\rho=\rho_0+\delta\rho$, $\theta=\theta_0+\delta\theta$, $\chi=\chi_0+\delta \chi$ and $\phi=\phi_0+\delta \phi$, where the subscript $0$ represents the mean field result and $\delta$ represents the fluctuation. $\delta \rho$, $\delta \chi$, and $\delta \phi$, which are massive due to repulsive interaction and spin-momentum locking induced by SOC respectively, are integrated out. The gapless phase fluctuation $\delta\theta$ is incorporated using imaginary time path integral,  $Z=\int D{\theta} e^{-\int d\tau dxdy \mathcal{L}(\theta)}$, where $\tau=it$, and $\mathcal{L}(\theta)$ is an effective low-energy Lagrangian. 

For $\Omega\ge\Omega_c$,  we obtain 
\begin{equation}
\mathcal{L}(\theta)=\alpha_\tau(\partial_\tau\theta)^2+\alpha({\bf \nabla}\theta)^2+\beta({\bf \nabla}^2\theta)^2\label{L}+\dots,
\end{equation}
where $\alpha_\tau=\frac{1}{4g_0}$, $\alpha=\frac{\rho_0}{2m}\frac{\Omega-\Omega_c}{\lambda+\Omega-\Omega_c}$, $\beta=\frac{\rho_0}{8m(\lambda+\Omega-\Omega_c)^2}$, and the ellipsis represents terms that are higher order in derivative expansion or contribute to observables only as higher order corrections (Supplementary Note 2). It is interesting to note that equation (\ref{L}) describes the Quantum Lifshitz model, an important tool for studying exotic phenomena in many subjects of modern physics, ranging from charge fractionalization and deconfinement transition in condensed matter physics \cite{Fradkin, Kivelson, dimer, Moessner1, Fradkin2, Ashvin} to quantum gravity in high energy physics\cite{Petr, Petr2, Holsheimer, Charmousis}. Whereas Quantum Lifshitz model has not been realized in a realistic system before, synthetic SOC naturally provides physicists an ideal simulator of it,  since all parameters in equation (\ref{L}) are well controlled. In particular, it allows one to access the Lifshitz point, where $\alpha ({\bf \nabla}\theta)^2$ vanishes at $\Omega_c$, by tuning $\Omega$. A sequence of exotic phenomena emerge here, such as the suppression of condensation, the rise of an algebraic bosonic liquid, and the deconfinement of topological excitations.  Unlike other systems where Quantum Lifshitz model remains a purely theoretical description, the field $\theta$ here directly corresponds to physical observables of ultra cold atoms, and all the above intriguing phenomena can be experimentally probed in our system.


{\bf Suppressed condensation and emerged algebraic quantum liquid}  Condensate density is provided by ${n_{0\uparrow}}=0$, ${n_{0\downarrow}}=n_{0}={\rho}e^{-\langle\theta^2\rangle}$. As $\langle\theta^2\rangle=Z^{-1}\int D{\theta}\theta^2 e^{-\int d\tau dxdy \mathcal{L}(\theta)}$, we have
\begin{equation}
n_{0}={\rho}_0\exp\left(-\frac{1}{4\pi\sqrt{\alpha_\tau}}\int dq \frac{q}{(\alpha q^2+\beta q^4)^{\frac{1}{2}}}\right).\label{n0}
\end{equation}
When $\alpha=0$, the sound velocity vanishes and the low-lying excitation spectrum becomes $\omega_{\bf q}=\sqrt{\beta/{\alpha_\tau}}q^2$. Such an unconventional collective excitation spectrum fundamentally changes thermodynamic properties of the system.  For instance, it leads to a linear specific heat at low temperatures, 
\begin{equation}
C_{\text v}=\frac{\pi}{12}\sqrt{\frac{2m\lambda^2}{\rho_0g_0}}T,
\end{equation}
 different from the conventional $T^2$ behavior in ordinary two-dimensional bosons. More importantly, one notes an infrared divergence $\int dq q^{-1} $  in equation (\ref{n0}), which usually occurs in one dimension for ground states of ordinary bosons. Such a divergence here destroys the two-dimensional condensation of interacting bosons at $T=0$.  This is a purely quantum effect, different from thermal fluctuation suppressed condensation in either two or three dimensions\cite{Zhou2, Sankar}. The characteristic long-range order at the ground state of ordinary two-dimensional systems is then replaced by an algebraic one. The one-body correlation function $\langle \hat{\Psi}_\downarrow^\dagger({\bf r})\hat{\Psi}_\downarrow(0)\rangle=\rho_0e^{-\langle(\theta(0)-\theta({\bf r}))^2 \rangle/2}$ becomes power-law like, 
\begin{equation}
\langle \hat{\Psi}_\downarrow^\dagger({\bf r})\hat{\Psi}_\downarrow(0)\rangle=\rho_0e^{-\frac{\gamma-\ln 2}{2K_c}}\left(\frac{|{\bf r}|}{\xi}\right)^{-\frac{1}{2K_c}}
\end{equation}
where  $\gamma$ is the Euler-Mascheroni constant, and $\xi=(2m\rho_0g_0)^{-1/2}$ is the healing length. $K_c=\pi\sqrt{\frac{\rho_0}{2g_0m\lambda^2}}$ is an effective Luttinger liquid parameter to characterize this algebraic quantum liquid. 

Whereas we focus on infinite systems in this manuscript, it is worth mentioning the finite size effects at the Liftshitz point.  In a finite system, a cutoff $\Lambda=1/L$ in the momentum space, where $L$ is the linear size of the system, removes the infrared divergence in the integral in  equation (\ref{n0}) even when $\alpha=0$, and therefore produces a finite-size-effect induced condensation, similar to an ordinary one-dimensional finite system. By choosing the lower bound of the integral as $\Lambda$, we obtain 
\begin{equation}
n_0=\rho_0\left(\frac{\xi}{L}\right)^{\frac{\lambda}{\pi\rho_0\xi}},
\end{equation}
which shows that $n_0$ decreases as a power-law function of the linear size of the system and eventually vanishes in the thermodynamic limit.

\begin{figure}[tbp]
\begin{center}
\includegraphics[width=3.6in]{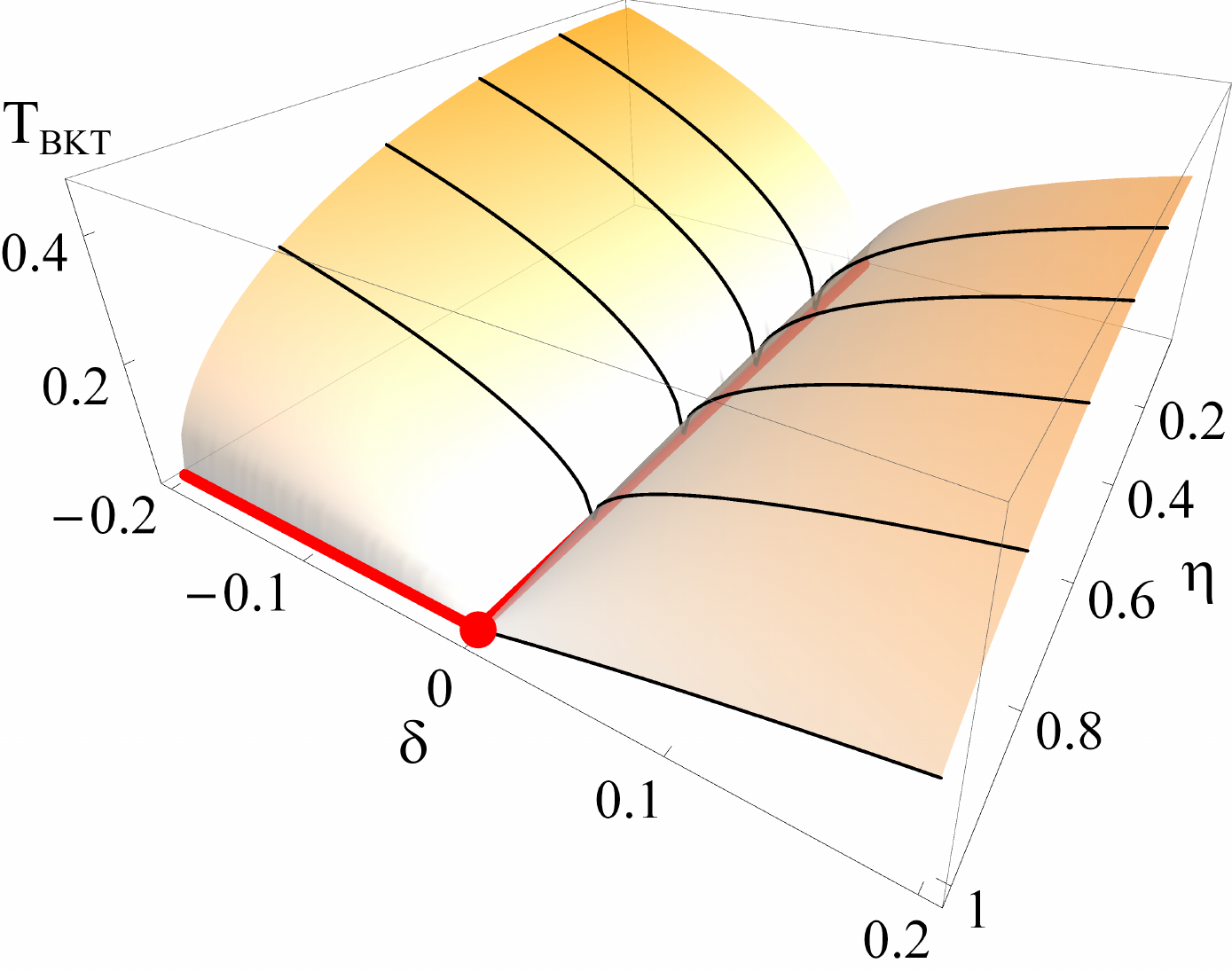}
\end{center}
\caption{BKT transition temperature $T_{BKT}$ as a function of $\delta$ and the anisotropy of SOC $\eta$. Black curves represent $T_{BKT}$ at different fixed values of $\eta$. Red lines represent where $T_{BKT}$ vanishes. For any values of $\eta<1$, $T_{BKT}$ is suppressed down to zero at one critical point $\delta_c(\eta)$. For the isotropic case $\eta=1$, $T_{BKT}$ remains zero if $\delta\le\delta_c(\eta=1)$. The red dot indicates the Lifshitz point.}
 \end{figure}

Near the critical point, whereas $n_0$ becomes finite, its amplitude is strongly suppressed due to the smallness of $\alpha$. We define $q^*$ that satisfies $\alpha q^{*2}=\beta q^{*4}$. As the quadratic and quartic terms are relatively more important in the Lagrangian for $q<q^*$ and $q>q^*$ respectively, the integral in equation (\ref{n0}) may be approximated by $\alpha^{-\frac{1}{2}}\int_0^{q^*} dq+ \beta^{-\frac{1}{2}}\int_{q^*}^\Lambda dq q^{-1}$, where $\Lambda=\xi^{-1}$ is a large momentum cutoff. Within this approximation, we obtain the scaling form for the condensate density near the critical point, 
\begin{equation}
{n_0}={\rho_0}e^{-\frac{1}{K}}\left(4{(\Omega-\Omega_c)(\lambda+\Omega-\Omega_c)}\xi^2\right)^{\frac{1}{2K}}\sim \delta^{\frac{1}{2K_c}} \label{ns},
\end{equation}
where $K={\lambda}K_c/({\lambda+\Omega-\Omega_c)}$ and $\delta=(\Omega-\Omega_c)/\Omega_c$. This result shows that synthetic SOC provides one a unique tool to control the condensation at the ground state without sophisticated designs of the microscopic Hamiltonian. Equation (\ref{ns}) is verified by an exact numerical evaluation of equation (\ref {n0}).  




When $\Omega<\Omega_c$, $\alpha$ becomes negative, and the phase gradient becomes finite. The low energy effective Lagrangian is reformulated around the new class of mean field solution, and we obtain
\begin{equation}\begin{split}
\mathcal{L}'(\theta)=&\tilde{\alpha}_\tau(\partial_\tau\theta)^2+\alpha_x( \partial_x\theta)^2+\beta_x(\partial_x^2\theta)^2\\
&~~~~~+\beta_y(\partial_y^2\theta)^2 + \beta_{xy} (\partial_x \partial_y \theta)^2+\dots
\label{L'},
\end{split}
\end{equation}
where the lengthy expressions for the coefficients are given in the Supplementary Note 2.  When $\Omega=\Omega_c$, $\mathcal{L}'(\theta)$ and $\mathcal{L}(\theta)$ become identical.  
If $\Omega<\Omega_c$,  $\mathcal{L}'(\theta)$ has only one quartic mode along the y direction. This come from the fact that $\epsilon_{\bf k}^{\pm}$ has an infinite degeneracy on the circle ${\bf k}=k_0$,  similar to the case where $\Omega=0$\cite{Zhou2, Jian}. A single quartic mode in two dimensions is not sufficient to destroy the long-range order at $T=0$, and the condensate fraction becomes finite again when $\Omega$ decreases from $\Omega_c$ , i.e., $n_{0}\sim (\Omega_c-\Omega)^{\frac{1}{2K_c}}$. 

The algebraic bosonic liquid at $\Omega=\Omega_c$ and the strong suppression of condensation near this Lifshitz point can be directly probed by measuring the momentum distribution and the correlation function.  When $\Omega=\Omega_c$, $n_\sigma({\bf q})=\int d {\bf r} e^{i{\bf q}\cdot{\bf r}}\langle \hat{\Psi}_\sigma^\dagger({\bf r})\hat{\Psi}_\sigma(0)\rangle\sim q^{\frac{1}{2K_c}-2}$. Such a power-law singularity signifies the algebraic order. In the vicinity of $\Omega_c$, the power-law like feature readily emerges in an intermediate momentum scale $q^*<q<\xi^{-1}$ due to the dominant quartic term in equation (\ref{L}) in this region.  This can be easily understood from the correlation function in the real space. In the region $\xi<|{\bf r}|< 1/q^{*}$, $\langle \hat{\Psi}^\dagger_\sigma ({\bf r})\hat{\Psi}_\sigma (0)\rangle$ is a power-law function, and approaches a constant $n_{0\sigma}$ when $|{\bf r}|\gg1/q^{*}$, as shown by the numerical results in figure 2. At the critical point, $1/q^{*}$ diverges and $\langle \hat{\Psi}_\sigma^\dagger({\bf r})\hat{\Psi}_\sigma(0)\rangle$ remains algebraic at arbitrarily large distances. Expressions for the correlation function at different length scales are given in the Supplementary Note 2.





{\bf Deconfinement transition and vanishing $T_{BKT}$}  We now turn to the deconfinement transition. For conventional two-dimensional bosons, vortices are the characteristic topological excitations and are confined by a logarithmic force $U_v=2\pi \rho_s\int d{\bf r}d{\bf r'}m({\bf r})\ln|({\bf r-r'}|)m({\bf r}')$, analogous to the Coulomb force in two-dimensional electrons\cite{PC}.  $m({\bf r})$ is the density of vortices at ${\bf r}$ and $\rho_s$ is the superfluid stiffness. A direct consequence of the confinement is a finite  Berezinsky-Kosterlitz-Thouless transition temperature $T_{BKT}$, below which free vortices are prohibited due to the binding of vortex and anti-vortex\cite{Thouless}. Synthetic SOC changes this fundamental property of two-dimensional bosons.  
 
At finite temperatures, the low-energy physics is dominated by the zero frequency mode $\omega=0$ of the Lagrangian in equation (\ref{L}). The effective Hamiltonian at $T\neq 0$ can be formulated as 
\begin{equation}
H_{eff}=\left \{
\begin{array}{cc}
\int d{\bf r}\alpha(\nabla\theta)^2, & \Omega>\Omega_c\\\\
\int d{\bf r}\beta(\nabla^2\theta)^2, & \Omega=\Omega_c\\\\
\int d{\bf r}\left(\alpha_x( \partial_x\theta)^2+\beta_y(\partial_y^2\theta)^2\right), &  \Omega<\Omega_c.
\end{array}
\right.
\end{equation}
For $\Omega>\Omega_c$, the Hamiltonian corresponds to an ordinary $XY$ model. $\rho_s=\alpha/2$ is controlled by $\delta=(\Omega-\Omega_c)/\Omega_c$. When $\alpha$ decreases down to zero, $\rho_s$ becomes zero, and the long-range Coulomb interaction among vortices disappears.  As $\nabla^2\theta=m(\bf r)$, one obtains $H_{eff}=\int d{\bf r}\beta(\nabla^2\theta)^2=\beta\int d{\bf r}d{\bf r'}m({\bf r})\delta^2({\bf r-r'})m({\bf r'})$ at $\Omega=\Omega_c$, i.e.,  the interaction between vortices becomes a short-range one\cite{dimer}. Once a vortex and anti-vortex pair is created by thermal excitations, the short-range interaction could not prevent them from deconfinement.  A direct consequence is then a vanishing $T_{BKT}$.  Near the critical point,  $T_{BKT}$ is given by the ordinary Berezinsky-Kosterlitz-Thouless theory, 
\begin{equation}
T_{BKT}={\pi}\alpha \sim \Omega-\Omega_c \label{BKT}.
\end{equation}




For $\Omega<\Omega_c$, the Hamiltonian can be regarded as an extreme case of the anisotropic $XY$ model 
\begin{equation}
\int d{\bf r}(\alpha_x( \partial_x\theta)^2+\alpha_y( \partial_y\theta)^2),\label{aXY}
\end{equation}
with $\alpha_y=0$. For $\alpha_y \neq 0$, one could perform a simple rescaling along the $y$ direction, and define ${\bf r'}=(x', y')=(x, (\alpha_x/\alpha_y)^{\frac{1}{2}}y)$ so that $H_{eff}'=\sqrt{\alpha_x\alpha_y} \int d{\bf r'}(\nabla_{\bf r'}\theta)^2$. This gives $T_{BKT}=\pi\sqrt{\alpha_x\alpha_y}$, and hence $T_{BKT}$ vanishes as $\alpha_y \rightarrow 0$, similar to a special case $\Omega=0$ studied before\cite{Jian}. $H_{eff}'$ shows that the logarithmic interaction between vortices vanishes if $\alpha_y=0$, and vortices are also deconfined, as a consequence of  the vanishing sound velocity  along the $y$ direction.    \\


$T_{BKT}$ can also be calculated for anisotropic SOC, where $0\le \eta<1$ and the Hamiltonian is an anisotropic XY model as in equation (\ref{aXY}). The coefficients of the effective Lagrangian are provided in the Supplementary Note 2.  There always exists a critical point $\Omega_c(\eta)$, where the coefficient of $(\partial_x \theta)^2$ vanishes. Due to the presence of $(\partial_y\theta)^2$ for $0\le \eta<1$, condensate fraction is  finite, with a minimum at $\Omega_c(\eta)$.  This fact can be qualitatively understood in the non-interacting limit, where the single particle spectrum becomes 
\begin{equation}
\epsilon_{\bf k}^{-}\approx\frac{1}{2m}\left(\frac{1}{4\lambda^2}(k_x^4+\eta^4k_y^4)+(1-\eta^2)k_y^2+\frac{\eta^2}{2\lambda^2}k_x^2k_y^2 \right)
\end{equation}
for small momenta. At $\Omega_c(\eta)$, the density of states becomes $\mathcal{N}(\epsilon)\sim \epsilon^{-1/4}$, which does not lead to divergent quantum depletion at zero temperature.  In contrast, $T_{BKT}$ is suppressed down to zero at $\Omega_c(\eta)$ due to the vanishing $(\partial_x\theta)^2$ term in the Hamiltonian, as shown in figure 3,  and vortices are deconfined at this point.  As mentioned before, the current Raman scheme corresponds to the extreme case where $\eta=0$. Experimentalists are readily able to observe the quenched $T_{BKT}$  in two dimensions. 

By taking images of  atomic densities,  the distribution of vortices, including their locations and separations,  have been measured\cite{Dalibard0, Shin}.  Moreover,  $T_{BKT}$  has been measured using a variety of schemes\cite{Bill, Dalibard2, Dalibard3}. This allows a direct visualization of the deconfinement transition in the system,  and a vanishing $T_{BKT}$ serves as a signature of such deconfinement. The scaling form of $T_{BKT}$ as shown in equation (\ref{BKT}) allows experimentalists to locate the deconfinement transition point without tuning $\Omega$ exactly at $\Omega_c$.


The study on two-dimensional bosons has been a long-term important effort in the field of ultra cold atom physics\cite{Bill, Dalibard2, Dalibard3}. Though the presence of a condensate as the ground state and a finite Berezinskii-Kosterlitz-Thouless transition temperature have been familiar to physicists,  we have shown that a synthetic spin-orbit coupling offers physicists a simple and practical scheme to defeat these standard textbook results by producing a quartic dispersion. Moreover, it leads to the realization of an ideal simulator of quantum Lifshitz model for accessing intriguing phenomena that are important in many other fields, such as deconfinement transitions of topological excitations. As it is of fundamental interest in condensed matter physics community to explore quantum phases without ordering at the ground state, we hope that our work will stimulate more studies on using the highly controllable synthetic SOC for taming the ordering in many-body states of ultra cold atoms and for exploring novel quantum phenomena that are not accessible in solids. We also hope that this atomic simulator of quantum Lifshitz model may be useful for high energy physics community on the topic of quantum gravity in the future.

QZ acknowledges useful discussions with C. Xu on Rokhsar-Kivelson Points in quantum dimer models. This work was supported by ECS/RGC 409513. This work was supported in part by the National Science Foundation under Grant No. PHYS-1066293, and a grant from the Simons Foundation. QZ  acknowledges the hospitality of the Aspen Center for Physics, where part of the manuscript were finished.


\begin{widetext} 

\noindent{\bf\Large Supplementary Materials}\\

\noindent{\bf\large Quartic dispersions in shaken lattices}
\vspace{0.1in}

By shaking a lattice, an effective spin-orbit coupling can be produced, where band indices play the role of spin[S1,S2]. If one shakes a one-dimensional lattice, it has been shown that a quartic dispersion could be produced[S1,S2,S3]. This method can be straightforwardly generalized to two dimensions. 

For a shaken square lattice, the lattice potential can be written as
\begin{equation}
V(x,y)=V_0\left(\cos\left ( \frac{2\pi}{d}(x+f\cos \omega t) \right)+\cos\left( \frac{2\pi}{d} (y+f\cos \omega t)\right)\right).
\end{equation}
When the shaking frequency $\omega$ is tuned to be close to the band gap $E_g$ between the $s$ and $p$ bands, these three bands are coupled by the photon-assistant hybridization. The Hamiltonian is written as
\begin{equation}
H_{shaken}=\left(\begin{array}{ccc}
-t_s(\cos k_x+\cos k_y)+\delta & C & C \\
C^* & -t_s\cos k_x+t_p\cos k_y & 0 \\
C^* & 0 & t_p\cos k_x-t_s\cos k_y
\end{array}\right)
\end{equation}
where $t_s$ and $t_p$ are the tunneling amplitude, $\delta=\omega-E_g$ is the detuning. The interband coupling $C$ can be written as
\begin{equation}
C=-\int \frac{V_0}{2}J_1(f) W_s(x,y)\cos\left (\frac{2\pi}{d} x \right)W_{p_x}(x,y)=-\int \frac{V_0}{2}J_1(f) W_s(x,y)\cos\left (\frac{2\pi}{d} y\right)W_{p_y}(x,y)
\end{equation}
where $J_1(f)$ is the Bessel function, and $W_s(x,y)$,  $W_{p_x}(x,y)$ and $W_{p_y}(x,y)$ are the Wannier wave functions of the three bands. 

Whereas this $3\times 3$ matrix can be exactly diagonized, it is useful to analytically demonstrate the emergence of quartic dispersion.  The energy of the dressed $s$ band can be written as

\begin{equation}
E_s=-t_s(\cos k_x+\cos k_y)+\delta-\frac{|C|^2}{(t_p+t_s)\cos k_y-\delta}-\frac{|C|^2}{(t_p+t_s)\cos k_x-\delta}
\end{equation}

It can be expanded as
\begin{equation}
E_s=\alpha(k_x^2+k_y^2)+\beta(k_x^4+k_y^4),
\end{equation}
where
\begin{equation}
\alpha=\frac{1}{2}\left(t_s-|C|^2\frac{t_s+t_p}{(t_s+t_p-\delta)^2}\right)
\end{equation}
\begin{equation}
\beta=\frac{1}{24}\left(-t_s-\frac{|C|^2(t_s+t_p)(\delta+5(t_s+t_p)))}{(t_s+t_p-\delta)^3}\right)
\end{equation}
A slight difference with the SOC results discussed in the main text is that the lattice potential reduces the symmetry to a four-fold one, instead of a full rotation symmetry in the momentum space. 

The quadratic term vanishes at a critical value $\delta_c$,
\begin{equation}
\delta_c^{\pm}=t_s+t_p\pm\left(\frac{t_s+t_p}{t_s}\right)^{\frac{1}{2}}|C|
\end{equation}
At $\delta_c^+$,
\begin{equation}
\beta=\frac{1}{4}\frac{t_s}{C}\sqrt{t_s(t_s+t_p)}>0
\end{equation}

\noindent{\bf  Effective theory for $0 < \eta \le 1$}
\vspace{0.1in}

We consider the two-dimensional Hamiltonian 
\begin{equation}\begin{split}
\mathcal H =& \int d^2 \vec r \left \{ \vec \Psi^{\dagger} \frac{1}{2m} \left[ \vec p^2 - 2 \lambda (\sigma_x p_x + \eta \sigma_y p_y) + 2 \lambda \Omega \sigma^z \right] \vec \Psi \right\}
 +  \int d^2 \vec r \left \{ \left( g_0 + \frac{g}{2} \right) [\vec \Psi^{\dagger}  \vec \Psi]^2 -  \frac{g}{2}   [\vec \Psi^{\dagger} \sigma_z \vec \Psi]^2  \right\},
\end{split}\end{equation}
with $0< \eta \le 1$ characterizing the spatial anisotropy in spin-orbit coupling. The system was analyzed with the ansatz
\begin{align}
\vec \Psi = \exp( i \sigma^y \pi/4) \sqrt{\rho} e^{ i \theta}
\left (
\begin{array}{r}
- \sin (\phi/2) \, e^{- i \chi /2} \\
\cos (\phi/2) \, e^{+ i \chi /2} 
\end{array}
\right).
\end{align}
Mean field solution that minimizes the energy functional is found, and it takes different forms for $\Omega$ larger than or smaller than its critical value $\Omega_c = \lambda - g_s m \rho_0 / \lambda$:
\begin{equation}\begin{split}\label{eq:}
\theta_0 =& \lambda \cos \phi_0 x; \\
\sin \phi_0 =& 
\left \{
\begin{array}{ccc}
\Omega / \Omega_c & \text{for} & \Omega < \Omega_c \\
1 & \text{for} & \Omega \geq \Omega_c 
\end{array}
\right. ;
\\
\chi_0 =& 0 .
\end{split}\end{equation}
The long-wavelength limit of the system is studied by developing Gaussian effective theories around the mean-field solution through incorporating fluctuations $\rho\rightarrow \rho_0 + \delta \rho$, $\phi \rightarrow \phi_0 + \delta  \phi$, $\chi \rightarrow \chi_0 + \delta  \chi$ and $\theta \rightarrow \theta_0 + \theta$.

\begin{enumerate}
\item $\Omega \geq \Omega_c$\\
Expanding the Lagrangian about the mean-field solution, up to quadratic order of the fluctuations we have
\begin{equation}\begin{split}\label{eq:}
\mathcal L \approx &   \delta  \rho \left[ g_0 - \frac{1}{8 m \rho_0} \vec \nabla^2\right]  \delta  \rho +  \delta  \rho \left [ i \partial_\tau \theta + \frac{\lambda}{2m}  \eta \partial_y \delta  \phi -  \frac{\lambda}{2m} \partial_x \delta  \chi \right]  + \delta  \phi \left[ \frac{\rho_0 \lambda^2}{2m} Z_{0} - \frac{\rho_0}{8m} \vec \nabla^2 \right] \delta  \phi \\
&+ \delta \phi \left[ - i \frac{\rho_0}{2} \partial_\tau \delta  \chi + \frac{\rho_0 \lambda}{m} \partial_x \theta \right]  
 +\delta  \chi\left[ \frac{\rho_0 \lambda^2}{2m} Z_{0}  -\frac{\rho_0}{8m} \vec \nabla^2\right] \delta  \chi + \delta  \chi \left[ \eta \frac{\lambda \rho_0}{m} \partial_y \theta \right]
 - \frac{\rho_0}{2m}\theta \vec \nabla^2 \theta, 
\end{split}\end{equation}
where
\begin{equation}\begin{split}\label{eq:}
Z_{0} = 1 + \frac{\Omega - \Omega_c}{\lambda} \geq 1
\end{split}\end{equation}
is defined in such a way that it is dimensionless and equals to $1$ at $\Omega = \Omega_c$.\\

The fields $\rho$, $\phi$ and $\chi$ are all gapped and can be integrated out, giving the effective Lagrangian for $\theta$ 
\begin{equation}\begin{split}\label{eq:}
\mathcal L'(\theta)= & 
\theta \left[ 
- \frac{1}{4 g_0} \partial_\tau^2
- \frac{\rho_0 (Z_0-1)}{2m Z_0} \partial_x^2  
- \frac{\rho_0 (Z_0 - \eta^2)}{2m Z_0} \partial_y^2  
+ \frac{\rho_0}{8 m \lambda^2 Z_0^2}   \partial_x^4   
+ \frac{(1+\eta^2) \rho_0}{8 m \lambda^2 Z_0^2} \partial_x^2 \partial_y^2  
+ \frac{\eta^2 \rho_0}{8 m \lambda^2 Z_0^2} \partial_y^4
\right.
\\&~~~~~~~
\left.
+\left( \frac{\lambda^2 Z_0 + g_0 m \rho_0}{8 g_0 \lambda^4  Z_0^3} + \frac{1-Z_0}{32 g_0^2 m \rho_0  Z_0 }\right) \partial_\tau^2 \partial_x^2
+ \left( \frac{\eta^2 (\lambda^2 Z_0 + g_0 m \rho_0)}{8 g_0 \lambda^4 Z_0^3} + \frac{\eta^2-Z_0 }{32 g_0^2 m \rho_0 Z_0}\right) \partial_\tau^2 \partial_y^2
\right] \theta. 
\end{split}\end{equation}

In particular, for the isotropic case with $\eta=1$ we obtain
\begin{equation}\begin{split}\label{eq:}
\mathcal L(\theta)= & 
\theta \left[ 
- \frac{1}{4 g_0} \partial_\tau^2
- \frac{\rho_0 (Z_0-1)}{2m Z_0} \vec \nabla^2  
+ \frac{\rho_0}{8 m \lambda^2 Z_0^2}  (\vec \nabla^2  )^2
+\left( \frac{\lambda^2 Z_0 + g_0 m \rho_0}{8 g_0 \lambda^4  Z_0^3} + \frac{1-Z_0}{32 g_0^2 m \rho_0  Z_0 }\right) \partial_\tau^2 \vec \nabla^2
\right] \theta.
\end{split}\end{equation}

Alternatively, at the critical field $\Omega = \Omega_c$, we have
\begin{equation}\begin{split}\label{eq:}
\mathcal L'(\theta)= & 
\theta \left[ 
- \frac{1}{4 g_0} \partial_\tau^2
- \frac{\rho_0 (1- \eta^2)}{2m } \partial_y^2  
+ \frac{\rho_0}{8 m \lambda^2 }   \partial_x^4   
+ \frac{(1+\eta^2) \rho_0}{8 m \lambda^2  } \partial_x^2 \partial_y^2  
+ \frac{\eta^2 \rho_0}{8 m \lambda^2  } \partial_y^4
\right.
\\&~~~~~~~
\left.
+\left( \frac{\lambda^2 + g_0 m \rho_0}{8 g_0 \lambda^4   }  \right) \partial_\tau^2 \partial_x^2
+ \left( \frac{\eta^2 (\lambda^2  + g_0 m \rho_0)}{8 g_0 \lambda^4  } + \frac{\eta^2-1 }{32 g_0^2 m \rho_0 }\right) \partial_\tau^2 \partial_y^2
\right] \theta, 
\end{split}\end{equation}
in which the $\sim \partial_x^2$ term is suppressed, leading to the vanishing of $T_{BKT}$.\\

At the quantum critical point $\Omega = \Omega_c$ and $\eta = 1$, we have $Z_0 = 1$ and the usual dominant spatial derivative $\sim \vec \nabla^2$ is suppressed, giving the Quantum Lifshitz model 
\begin{equation}\begin{split}\label{eq:}
\mathcal L_c(\theta)= & 
\theta \left[ 
- \frac{1}{4 g_0}   \partial_\tau^2
+ \frac{\rho_0}{8 m \lambda^2}  (\vec \nabla^2  )^2
\right] \theta,
\end{split}\end{equation}
where the term $\sim \partial_\tau^2 \vec \nabla^2$  can be dropped as it now enters as a higher order correction in the absence of the usual $\sim \vec \nabla^2$ term.

\item $\Omega < \Omega_c$\\
In the same manner the Lagrangian is expanded as
\begin{equation}\begin{split}
\mathcal L \approx &
\delta  \rho \left[ \tilde g_0 - \frac{1}{8 m \rho_0} \vec \nabla^2\right]\delta \rho 
+ \delta  \rho \left[ i \partial_\tau \theta + \frac{i}{2} \cos\phi_0 \partial_\tau \delta  \chi - (g_s \rho_0 \cos\phi_0 \sin\phi_0 ) \delta  \phi  + \frac{\lambda}{2m} \eta \partial_y \delta  \phi - \frac{\lambda}{2m}   \sin^2 \phi_0 \partial_x \delta  \chi  \right]\\
&+ \delta  \phi \left[ \left(\frac{\lambda^2 \rho_0}{2m} - \frac{g_s \rho_0^2}{2} \cos^2 \phi_0 \right)  - \frac{\rho_0}{8m} \vec \nabla^2 \right] \delta  \phi 
+ \delta  \phi \left[ - \frac{i}{2} \rho_0 \sin\phi_0 \partial_\tau \delta \chi -\frac{\lambda\rho_0}{2m}  \sin\phi_0 \cos\phi_0 \partial_x \delta  \chi + \frac{\lambda \rho_0}{m} \sin \phi_0 \partial_x \theta \right] \\
& +\delta  \chi \left[ \frac{\lambda^2 \rho_0}{2m}\sin^2 \phi_0 - \frac{\rho_0}{8m} \vec \nabla^2 \right] \delta  \chi + \delta  \chi \left[ \frac{\eta \lambda \rho_0}{m} \sin\phi_0 \partial_y \theta - \frac{\rho_0 }{2 m} \cos\phi_0 \vec \nabla^2 \theta \right]\\
& - \frac{\rho_0}{2m}\theta \vec \nabla^2 \theta
\end{split}\end{equation}
where $\tilde g_0 = g_0 + g_s \cos^2 \phi_0/2$. Note the additional dependence on $\phi_0$ as it is now a function of $\Omega$.\\

In integrating out each of the massive fields $\rho$, $\phi$ and $\chi$, the coefficients in the effective Lagrangian receive corrections and the expressions are substantially more complicated than those for $\Omega \geq \Omega_c$. The corrections are expressed in terms of symbols $Z's$ and $Y's$, which are all defined in such a way that they are dimensionless and  equal to $1$ at $\Omega = \Omega_c$ (regardless of the value of $\eta$). The effective Lagrangian for $\theta$ is 
\begin{equation}\begin{split}\label{eq:}
\mathcal L'(\theta) =&
 \theta \left[ 
- \frac{1}{4  g_0} Z_{\tau \tau}^{  \theta}  \partial_\tau^2   
- \frac{\rho_0}{2m} (1 - Z_{xx}^{  \theta}) \partial_x^2
 - \frac{\rho_0 (1-\eta^2)}{2m}  \partial_y^2  
 + i \frac{\rho_0}{2 \lambda} (1 - Z_{\tau x}^{  \theta}) \partial_\tau \partial_x 
 \right] \theta \\
&
+\theta \left [
 \frac{\lambda^2 + g_0 m \rho_0}{8 g_0 \lambda^4} Z_{\tau\tau x x}^{\theta}
   \, \partial_\tau^2  \partial_x^2
+\left( \frac{\eta^2 (\lambda^2 + g_0 m \rho_0)}{8 g_0 \lambda^4} + \frac{\eta^2 - 1}{32 g_0^2 m \rho_0}\right) Z_{\tau\tau y y}^{\theta}
 \partial_\tau^2  \partial_y^2\right.\\
&~~~~~~~\left.
-i \frac{1}{  g_0 m \lambda  } (1-Z_{\tau x y y}^{ \theta}) \, \partial_\tau \partial_x \partial_y^2
-i \frac{1}{  g_0 m \lambda  } (1- Z_{\tau x x x}^{ \theta})\, \partial_\tau \partial_x^3 \right.\\\
&~~~~~~~\left.
+ \frac{  \rho_0  }{8 m \lambda^2} Z_{xxxx}^{\theta } \partial_x^4
+ \frac{(1+\eta^2) \rho_0}{8 m \lambda^2}  Z_{xxyy}^{\theta} \partial_x^2 \partial_y^2
+ \frac{\eta^2 \rho_0}{8 m \lambda^2} Z_{yyyy}^{\theta } \partial_y^4
 \right] \theta,
\end{split}\end{equation}
and the lengthy expressions for $Z$ and $Y$ in terms of the system parameters are listed in the last section of this supplementary material. Note also that the coefficients of the effective Lagrangian are continuous at $\Omega = \Omega_c$.

\end{enumerate}

\noindent{\bf Condensate fraction and correlation functions}
\vspace{0.1in}

The effective Lagrangian was computed via derivative expansion and terms up to $4^{\text{th}}$ order are kept. The most general effective Lagrangian considered can be parameterize as 
\begin{equation}\begin{split}\label{eq:}
\mathcal L (\theta) =& \alpha_\tau (\partial_\tau \theta)^2 + \alpha_x  (\partial_x \theta)^2 + \alpha_y (\partial_y \theta)^2  + \beta_x (\partial_x^2 \theta)^2+ \beta_y (\partial_y^2 \theta)^2+ \beta_{xy} (\partial_x \partial_y \theta)^2+ \\
&+i \gamma_x (\partial_\tau \theta) (\partial_x \theta)+ \zeta_x (\partial_\tau \partial_x \theta)^2 + \zeta_y (\partial_\tau \partial_y\theta)^2 + i \epsilon_{xxx} (\partial_\tau \partial_x \theta) (\partial_x^2 \theta)+ i \epsilon_{xyy} (\partial_\tau \partial_x \theta) (\partial_y^2 \theta)
\end{split}\end{equation}
where terms like $\sim (\partial_\tau \theta)(\partial_y\theta)$, which can contribute to the observables to the same order of our calculations, are not generated in integrating out the heavy modes and are therefore absent.\\

The condensate density, given by $n_0 = n_{0\uparrow} + n_{0\downarrow} = \rho_0 e^{- \langle \theta^2\rangle}$, is found by numerically evaluating the integral
\begin{equation}\begin{split}\label{eq:}
\langle \theta^2 \rangle =
\int \frac{d \omega d^2 \vec q}{(2 \pi)^3} \frac{1}{A \omega^2 +  i B \omega + C} = \frac{1}{8 \pi^2 } \int d^2 \vec  q \frac{1}{\sqrt{AC + B^2/4}}
\end{split}\end{equation}
where
\begin{equation}\begin{split}\label{eq:}
A =& \alpha_\tau + \zeta_x \, q_x^2 + \zeta_y \,  q_y^2\\
B=& \gamma_x q_x +  \epsilon_{xyy} q_x q_y^2 + \epsilon_{xxx} q_x^3\\
C=& \alpha_x \, q_x^2 + \alpha_y\, q_y^2 + \beta_x \,q_x^4 + \beta_y \,q_y^4 + \beta_{xy}\,q_x^2 q_y^2\\
\Rightarrow AC+B^2/4 \approx& \left( \alpha_\tau \alpha_x + \gamma_x^2/4\right) q_x^2 + (\alpha_\tau \alpha_y )q_y^2 
+ \left( \alpha_\tau \beta_x + \alpha_x \zeta_x + \gamma_x \epsilon_{xxx}/2\right) q_x^4\\
&~~~ + (\alpha_\tau \beta_y + \alpha_y \zeta_y) q_y^4 + \left( \alpha_\tau \beta_{xy} + \alpha_x \zeta_y + \alpha_y \zeta_x + \gamma_x \epsilon_{xyy}/2\right) q_x^2 q_y^2\\
=&  \alpha_\tau \left( \tilde \alpha_x q_x^2 + \tilde \alpha_y q_y^2 + \tilde \beta_x q_x^4 + \tilde \beta_y q_y^4 + \tilde \beta_{xy} q_x^2 q_y^2 \right)
\end{split}\end{equation}

Similarly, the correlation function $\left\langle \vec{\Psi}^\dagger (\vec r) \vec{\Psi}(\vec 0 )\right\rangle = \rho_0 e^{- \langle \left( \theta(\vec r) - \theta(\vec 0) \right)^2\rangle/2}$ is found by numerically evaluating 
\begin{equation}\begin{split}\label{eq:}
\left \langle \left(\theta(\vec r ) - \theta(\vec0)\right)^2 \right\rangle 
= \frac{1}{8 \pi^2 } \int d^2 \vec  q \frac{1 - \cos{(\vec q \cdot \vec r)}}{\sqrt{AC + B^2/4}}.
\end{split}\end{equation}
The cutoff of the momentum integrals are set by the healing length: $q_{max} = \xi^{-1}$.\\

For the isotropic case with $\eta =1 $ and $\Omega > \Omega_c$, we have $\tilde \alpha_x = \tilde \alpha_y = \tilde \alpha$ and $\tilde \beta_x = \tilde \beta_y = \tilde \beta_{xy}/2 = \tilde \beta$. The characteristic momentum scale is given by $q^* = \sqrt{\tilde \alpha / \tilde \beta}$ and the integral is given by 
\begin{equation}\begin{split}\label{eq:}
\left \langle \left(\theta(\vec r ) - \theta(\vec0)\right)^2 \right\rangle 
=&  \frac{1}{4 \pi \sqrt{\alpha_\tau \tilde \beta} } \int_{0}^{(q ^*\xi)^{-1}}  d \tilde q  \frac{1 - J_0{(\tilde q \tilde r )}}{\sqrt{1+ \tilde q^2}},
\end{split}\end{equation}
where $\tilde q$ and $\tilde r$ are both dimensionless (measured in units of $q*$ and $(q^*)^{-1}$). For $| \vec r| \gg (q^*)^{-1}$, the integral is dominated by the small momentum contribution and so
\begin{equation}\begin{split}\label{eq:}
\left\langle \vec{\Psi}^\dagger (\vec r) \vec{\Psi}(\vec 0 )\right\rangle 
\approx&  \rho_0 \exp{\left(- \frac{1}{8 \pi \sqrt{\alpha_\tau \tilde \alpha} } \left( \frac{1}{\xi} - \frac{1}{r}\right) \right)}  \\
\rightarrow &\rho_0 \exp{\left(- \frac{1}{8 \pi \xi \sqrt{\alpha_\tau \tilde \alpha} }  \right)}   \text{ as } r \rightarrow \infty,
\end{split}\end{equation}
giving the standard long-range correlation in the condensate.\\

Alternatively, for intermediate values of $|\vec r|$ such that $\xi \ll |\vec r | \ll (q^*)^{-1}$, the large momentum contribution of the integral dominates and it gives a power-law correlation function 
\begin{equation}\begin{split}\label{eq:}
\left\langle \vec{\Psi}^\dagger (\vec r) \vec{\Psi}(\vec 0 )\right\rangle 
\sim&  \rho_0 \left( \frac{| \vec r|}{\xi}\right)^{\frac{1}{2 K}},
\end{split}\end{equation}
where $K = 4 \pi \sqrt{\alpha_\tau \tilde \beta}$ is the effective Luttinger liquid parameter. As such the correlation function has a crossover behavior set by the characteristic length scale $(q^*)^{-1}$. In the limit $\Omega \rightarrow \Omega_c$, $(q^*)^{-1}$ diverges and therefore the correlation function is dominated by the power-law behavior.
\vspace{0.1in}

\noindent{\bf Expressions for $Z$'s and $Y$'s in the effective Lagrangian for $\Omega < \Omega_c$} 
\vspace{0.1in}

\begin{equation*}\begin{split}\label{eq:}
Z_{\tau\tau}^{ \theta} =& \frac{g_0}{\tilde g_0} + \frac{g_s^2 g_0 m \rho_0 \sin^2 \phi_0}{2 \tilde g_0^4 \lambda^2  Z_0^\phi}\cos^2\phi_0; ~~~~~~~~~~~~
Z_{xx}^{  \theta}= \frac{\sin^2\phi_0}{Z_0^\phi}; ~~~~~~~~~~~~
Z_{\tau x}^{  \theta} =  1 - \frac{g_s\sin^2\phi_0}{\tilde g_0 Z_0^\phi} \cos \phi_0;\\
Z_{\tau\tau x x}^{\theta} =&
\frac{ (Y_{\tau x}^{\chi \theta} )^2 (\lambda^2 + 2 g_0 m \rho_0)^2 - Y_{\tau \tau x x}^{\theta \theta} \sin^2\phi_0 \lambda^4 -4 g_0 m \rho_0 \lambda^2 (1- Y_{\tau \tau}^{\chi \theta} )  (1 - Y_{xx}^{\chi \theta})  }{4 g_0 m \rho_0 (\lambda^2 + g_0 m \rho_0) \sin^2 \phi_0 };\\
Z_{\tau \tau y y}^{\theta} =& 
\frac{4 \eta^2  g_0^2 m^2 \rho_0^2 Z_{\tau \tau}^{\chi} + 4 g_0 m \rho_0 \lambda^2 [  (Y_{\tau \tau}^{\chi \theta} -1 )\cos \phi_0 + \eta^2 Y_{\tau \tau y}^{\chi \theta}  \sin \phi_0  ] - Y_{\tau \tau y y }^{\theta \theta} \sin^2 \phi_0 \lambda^4}{ (\eta^2 (\lambda^2 + 2 g_0 m \rho_0)^2 - \lambda^4)\sin^2\phi_0};\\
Z_{\tau x y y}^{ \theta}=& 1 - \frac{  (\lambda^2 + 2 g_0 m \rho_0) Y_{\tau x}^{\chi \theta} \cos\phi_0 + 2 \lambda^2  [4 \eta^2 (Z_{\tau x}^{\chi  } -1) + \eta^2 (1 - Y_{\tau x y}^{\chi \theta} )   \sin \phi_0 +4 (Y_{\tau x y y }^{\theta \theta} -1) \sin^2 \phi_0   ] }{8  \lambda^2 \sin^2 \phi_0};\\
Z_{\tau x x x }^\theta =&  1- \frac{ (\lambda^2 + 2 g_0 m \rho_0) (1-Y_{xx}^{\chi \theta} ) Y_{\tau x}^{\chi \theta} +8 \lambda^2(Y_{\tau xxx}^{\theta \theta} -1)  \sin^2\phi_0  }{8  \lambda^2 \sin^2 \phi_0};\\
Z_{xxxx}^\theta =& Y_{xxxx}^{\theta \theta} - \frac{(Y_{xx}^{\chi \theta}-1)^2 }{\sin^2 \phi_0}; ~~~~~~~~~~~~
 Z_{yyyy}^{\theta }=  \frac{\eta^2 - \cos^2\phi_0}{\eta^2 \sin^2 \phi_0};\\
 Z_{xxyy}^{\theta} =&
 \frac{ Y_{xxyy}^{\theta \theta} \sin^2 \phi_0 + Z_{xx}^{\chi} \eta^2  }{   (1+\eta^2) \sin^2 \phi_0}
 + \frac{  4 g_0 m \rho_0 (Y_{xx}^{\chi \theta} -1) \cos \phi_0 - \eta^2 \lambda^2 (Z_{xx}^{\chi} + Y_{xxyy}^{\theta \theta} \sin^2 \phi_0 - 2 Y_{xxy}^{\chi \theta} \sin \phi_0) }{2 g_0 m \rho_0 (1+\eta^2) \sin^2 \phi_0} ;\\
 Z_{\tau \tau}^{\chi} = & \frac{g_0^2\sin^2\phi_0}{\tilde g_0^2 Z_0^\phi} + \frac{\lambda^2}{2 \tilde g_0 m \rho_0}\cos^2 \phi_0;\\
 Z_{xx}^{\chi}=& \frac{g_0 (2 \tilde g_0 m \rho_0 - \lambda^2 \sin ^4 \phi_0)}{\tilde g_0 (2 g_0 m \rho_0 - \lambda^2)} - \frac{g_0 (g_s+2g_0)^2 m \rho_0  \sin^2\phi_0}{2 \tilde g_0^2 Z_0^\phi(2 g_0 m \rho_0 - \lambda^2) } \cos^2\phi_0;\\
 Z_{\tau x}^{\chi}=&   1 - \frac{g_0 \tilde g_0 \lambda^2 Z_0^\phi - (g_s+ 2 g_0) g_0^2 m \rho_0 }{8 \tilde g_0^2   \lambda^2 Z_0^\phi}\sin^2\phi_0 \cos \phi_0 ;\\
 Y_{\tau\tau}^{\chi \theta} =& 1 - \frac{g_0 \tilde g_0 \lambda^2 Z_0^\phi - g_sg_0^2 m \rho_0 \sin^2\phi_0}{ \tilde g_0^2  \lambda^2 Z_0^\phi}\cos \phi_0;\\
Y_{xx}^{\chi \theta} =&  1  - \frac{  2 \tilde g_0 Z_0^\phi + (g_s +2g_0) \sin^2\phi_0 }{2 \tilde g_0   Z_0^\phi} \cos \phi_0;\\
Y_{\tau x}^{\chi \theta} =& \frac{g_0 (Z_0 ^\phi \lambda^2 + 2\tilde g_0 m \rho_0) \sin^2\phi_0}{\tilde g_0 (\lambda^2 + 2 g_0 m \rho_0) Z_0^\phi } + \frac{g_s^2 g_0 m \rho_0 \sin^4 \phi_0}{2 \tilde g_0^2 (\lambda^2 + 2 g_0 m \rho_0) Z_0^\phi} \cos^2\phi_0;\\
 \end{split}\end{equation*}
 
 \begin{equation*}\begin{split}\label{eq2:}
Y_{\tau \tau y}^{\chi \theta}=& \frac{g_0 \sin\phi_0}{\tilde g_0 Z_0^\phi}; ~~~~~~~~~~~~
Y_{\tau x y}^{\chi \theta} = 1 -\frac{g_0}{ \tilde g_0 Z_0^\phi}\sin\phi_0 \cos \phi_0; ~~~~~~~~~~~~
Y_{xxy}^{\chi \theta}=   \frac{g_0 \sin^3\phi_0}{\tilde g_0 Z_0^\phi};\\
Y_{\tau\tau x x}^{\theta \theta} =&  \frac{g_0^2}{\tilde g_0^2} + \frac{g_s^2 g_0^2 m \rho_0 (Z_0^\phi \lambda^2 + \rho_0 \tilde g_0 m Z_2^\phi) \sin^2 \phi_0}{\tilde g_0^3 \lambda^4 (Z_0^\phi)^2} \cos^2\phi_0;\\
Y_{\tau\tau y y}^{\theta \theta} =& \frac{g_0^2 ( Z_0^\phi - \eta^2)}{\tilde g_0^2 Z_0^\phi} + \frac{g_s^2 g_0^2 m \rho_0 ( 2   \lambda^2 Z_0^\phi - \eta^2 \lambda^2 + 2 \tilde g_0 m \rho_0  Z_2^\phi ) \sin^2 \phi_0}{2 \tilde g_0^3 \lambda^4  (Z_0^\phi)^2} \cos^2\phi_0;\\
Y_{\tau x y y}^{\theta\theta}=&  1- \frac{g_sg_0 (\lambda^2 (Z_0^\phi  -\eta^2) + 2 \tilde g_0 m \rho_0 Z_2^\phi) \sin^2 \phi_0}{16 \tilde g_0^2 \lambda^2 (Z_0^\phi)^2} \cos \phi_0;\\
Y_{\tau x x x}^{\theta\theta}=&  1 - \frac{g_sg_0 (\lambda^2 Z_0^\phi  + 2 \tilde g_0 m \rho_0 Z_2^\phi ) \sin^2\phi_0}{16 \tilde g_0 \lambda^2 (Z_0^\phi)^2} \cos \phi_0
;\\
Y_{xxyy}^{\theta \theta} =& \frac{g_0 (2 \tilde g_0 m \rho_0 Z_2^\phi - \eta^2 \lambda^2) \sin^2\phi_0 }{\tilde g_0 (2 g_0 m \rho_0 - \eta^2 \lambda^2) (Z_0^\phi)^2} ;\\
Y_{xxxx}^{\theta \theta} =&  \frac{Z_2^\phi \sin^2\phi_0}{(Z_0^\phi)^2};\\
 Z^{\phi}_{0}=&  1 - \frac{m  g_s\rho_0 (g_s+2 g_0)}{2 \lambda^2\tilde g_0 } \cos^2 \phi_0 ; ~~~~~~~~~~~~
 Z^{\phi}_{2}=   1 + \frac{g_s^2   \sin^2\phi_0}{4   \tilde g_0 ^2 }  \cos^2\phi_0 .
\end{split}\end{equation*}

\vspace{0.1in}

{\bf\large Supplementary References}
\vspace{0.1in}

[S1] Zhang, S.L., and Zhou, Q.,  Shaping topological properties of the band structures in a shaken optical lattice, arXiv:1403.0210 

[S2] Zheng, W., and Zhai, H., Floquet Topological States in Shaking Optical Lattices, Phys. Rev. A 89, 061603 (2014) 

[S3]  Parker, C. V., Ha, L. , Chin, C., Direct observation of effective ferromagnetic domains of cold atoms in a shaken optical lattice, Nature Physics 9, 769-774 (2013)

\end{widetext}

\end{document}